\documentclass[runningheads]{llncs}

\usepackage{amsfonts}
\usepackage{amsmath,amssymb,enumitem}
\usepackage{bm}
\usepackage{algorithm}
\usepackage{algorithmic}
\usepackage{float}           
\usepackage{color}
\usepackage{booktabs} 
\usepackage{ifthen,latexsym,syntonly}
\usepackage{rotating}
\usepackage{lscape}
\usepackage{graphicx}
\usepackage{color}
\usepackage{todonotes}
\usepackage[hypertexnames=false,hyperfootnotes=false]{hyperref}
\usepackage{cleveref}
\usepackage{url}
\usepackage{xcolor}
\usepackage{placeins}
\usepackage{float}
\usepackage{booktabs,adjustbox}
\newtheorem{assumption}{Assumption}

\usepackage[normalem]{ulem}
\newcommand{\defeq}{\ensuremath{\triangleq}}

\providecommand{\U}[1]{\protect\rule{.1in}{.1in}}

\makeatletter
\renewcommand{\inst}[1]{\ensuremath{^{#1}}}
\makeatother

\usepackage{ifthen}

\provideboolean{submissionversion}

\ifthenelse{\boolean{submissionversion}}{
  \renewcommand{\todo}[2][1]{}
}{
}

\provideboolean{showedits}
\setboolean{showedits}{true}

\ifthenelse{\boolean{showedits}}{
  
  \newcommand{\soutred}[1]{\textcolor{red}{\sout{#1}}}
}{
  
  \newcommand{\soutred}[1]{}
}

\provideboolean{shownewedits}
\setboolean{shownewedits}{true}

\ifthenelse{\boolean{shownewedits}}{
  
  \newcommand{\newsout}[1]{\textcolor{orange}{\sout{#1}}}
}{
  
  \newcommand{\newsout}[1]{}
}

\title{Quantifying Sub-Optimality in Routing for Automated Market Makers}
\titlerunning{Quantifying Sub-Optimality in Routing for Automated Market Makers}

\author{
  Weiye Xi\inst{1} \and
  Ciamac C. Moallemi\inst{1}
}
\institute{
  \inst{1} Decision, Risk, and Operations Division,\\
  Graduate School of Business, Columbia University
}

\begin{document}
\maketitle

\vspace{-0.5cm}
\begin{abstract}
We provide a large-scale empirical audit of DEX routing using 2.98 million WETH–USDC swaps on Ethereum. Comparing realized routes with optimized benchmarks, we measure an average shortfall of 2.02 bps per trade or \$24 million. To attribute losses, we introduce three reproducible optimal benchmarks: a Support-Constrained Optimum (SCO) that evaluates split quality conditional on the pools actually used; a Full-Venue Optimum (FVO) that considers all available pools to quantify the value of broader pool access; and a Gas-Aware FVO (G-FVO) that augments FVO with gas costs to capture the trade-off between additional pool usage and gas expenditure. Computing these benchmarks at scale is enabled by a bisection-based algorithm for optimal routing across multiple pools for the same token pair. Two regularities emerge. First, information timeliness is crucial: moving from execution-time state to one-block lagged state optimization significantly raises mean shortfall and additional delays further degrade performance, albeit with diminishing increments; evaluated on the same stale snapshots, realized routes lie closer to optimal, indicating timing-mismatch as a key component. Second, inefficiency is heterogeneous and heavy-tailed: small trades suffer higher percentage losses, while a few extreme outliers dominate the aggregate dollar shortfalls. Finally, we demonstrate that sandwiching attacks drive a significant fraction of routing sub-optimality. Our benchmark protocol and algorithm offer a rigorous, reproducible basis for evaluating and improving information-timely, gas-aware routing.

\end{abstract}

\section{Introduction}
\label{sec:intro}
Decentralized exchange (DEX) aggregators rely on smart routers to split trades among multiple automated market maker (AMM) pools for the best execution. Today’s routers must decide where to split flow across many concurrent pools and when to execute under gas costs and rapidly changing pool states. Despite an active design space, the field lacks a careful, large‑scale measurement of a basic question: \emph{How optimal are current routing decisions in practice, relative to economically motivated objectives?}

We address this question by conducting a comprehensive empirical audit of on-chain routing performance, using 2.98 million WETH–USDC swaps on Ethereum mainnet. We quantify the output value lost due to suboptimal routing by comparing each realized trade to optimized benchmarks. On average, current routing strategies sacrifice approximately 2.02 bps of potential output per trade, which aggregates to \$24 million of forgone value across roughly \$120 billion of volume. 

To rigorously evaluate routing suboptimality, we design three optimal routing benchmarks to assess each trade’s potential performance. First, the \emph{Support-Constrained Optimum} (SCO) benchmark restricts the pool set to only those actually used by the trade, measuring the quality of the volume split across those pools. Second, the Full-Venue Optimum (FVO) benchmark considers all available pools without accounting for gas costs, revealing the potential gain from accessing a broader liquidity pool set. Finally, the Gas-Aware Full-Venue Optimum (G-FVO) benchmark extends FVO by accounting for gas costs, capturing the trade-off between using additional pools and incurring extra gas fees. This suite of benchmarks allows us not only to measure the suboptimality of current routing decisions but also to decompose these inefficiencies and measure both \emph{the shortfall from failing to utilize the optimal subset of pools} and \emph{the cost of imperfect flow-splitting across the pools that were utilized}.

In order to efficiently compute our benchmarks, we describe an efficient algorithm for optimally routing across multiple pools for the same token pair based on bisection. Our method also accounts for discrete gas costs when selecting pools. Our method finds the exact maximum-output allocation, and we prove its correctness and characterize its complexity. We use this routine to construct our optimized benchmarks.

We investigate the value of timeliness of information in routing. Using a controlled experiment on information staleness, we isolate the effect of pool-state freshness on routing outcomes. Optimizing trades on slightly stale information significantly worsens outcomes compared to optimizing using up-to-date information, by about 1.29 bps under FVO (approximately \$15.5 million in our sample) and 1.78 bps under G-FVO (\$21.4 million). Further delays continue to erode performance, albeit by diminishing increments. Conversely, when we evaluate the realized routes using earlier (pre-execution) snapshots, they appear less suboptimal. This indicates that a sizeable portion of suboptimality is due to timing mismatch: \emph{Routes are planned using outdated information relative to execution.}

We further observe significant heterogeneity in routing inefficiency across trades, with a heavy-tailed loss distribution. Small swaps ($<\$1{,}000$) tend to incur disproportionately larger relative losses (in bps) than larger trades ($>\$10{,}000$). Large trades, while nearly optimal in percentage terms, can still suffer greater absolute slippage. Moreover, the suboptimality distribution has a long right tail driven by a few extreme outliers, consistent with potential adversarial execution (e.g., sandwich attacks) or transient liquidity shocks. These findings underscore the presence of heavy-tailed risk: most trades incur minimal loss, but rare worst-case events can result in much larger inefficiencies.

Finally, we quantify the contribution of sandwich attacks to measured routing inefficiency. We tag victim transactions whenever at least one of the four WETH-USDC pools is sandwiched and compare realized routes to the \emph{SCO}/\emph{FVO}/\emph{G-FVO} optima at execution\mbox{-}time snapshots. Sandwiched trades exhibit a pronounced upward shift in both the level and dispersion of suboptimality, with heavier right tails that persist even under gas\mbox{-}aware evaluation. Cross\mbox{-}router differences are consistent with execution context: exposure concentrates in publicly submitted flow (e.g., Universal Router), whereas solver\mbox{-}mediated batch auctions and private order flow (e.g., CoWSwap) show negligible incidence. These patterns reinforce our timing results: when planning on stale state and facing reordering, timing mismatch widens the execution\mbox{-}time gap. They also provide a concrete channel through which information staleness and pool activation interact to generate heavy\mbox{-}tailed losses.

In summary, our study provides a detailed empirical assessment of DEX aggregator routing quality and pinpoints multiple sources of suboptimality in current practice. Together, these contributions demonstrate economically significant inefficiencies in existing router outcomes and offer new tools and insights to guide the design of more optimal, timing-aware aggregation strategies.

\section{Related Literature}
Our work relates to several strands of research:
(1) \emph{Routing and aggregation.}
Routers and aggregators split flow across venues facing curved liquidity and heterogeneous fees. On the theory side, our work builds on the formulation of optimal routing problems as convex programs  \cite{AngerisEtAl2022CFMMRouting,diamandis2023efficient}. Recent work extends these formulations to richer execution environments (e.g., hooks) and studies their implications for route design \cite{Chitra2025Hooks}. On the empirical side, early evidence documents routing suboptimality and its drivers \cite{yaish2023suboptimality,kulkarni2023routing}. Relative to this literature, we introduce a gas-aware activation layer and algorithms with guarantees, and deliver an implementable benchmark suite (SCO/FVO/G-FVO).
(2) \emph{Execution frictions, MEV, and order‑submission context.}
The sensitivity of realized outcomes to timely state connects to the literature on mempools, reordering, and extractable value  \cite{Daian2020FlashBoys,Qin2022DarkForest}. Differences across routers in our staleness experiments are consistent with institutional features, e.g., solver‑mediated batch auctions and private order flow in CoW Protocol \cite{CoWProtocol2023Whitepaper,CoWProtocol2023Docs} that can mitigate within‑block state sensitivity.

\section{Optimal Routing Methodology}
\label{sec:routing}

\subsection{Routing Problem Formulation}
\label{subsection:Routing Problem Formulation}

Consider the following setup. An agent seeks to \emph{spend a fixed total amount of the input token} and receive as much of the output token as possible by splitting order flow across a collection of AMM pools indexed by $j\in\mathcal{J}=\{0,1,\ldots,n-1\}$. Each AMM pool $j$ is a two–asset pool that supports swaps between \emph{token~0} and \emph{token~1} (e.g., WETH and USDC). Let $q_j\ge 0$ denote the input routed to pool $j$, and let $Q>0$ be the \emph{total input budget}.

Note that here we are focusing on a simplified version of the full routing problem: we restrict attention to a setting in which all pools under consideration are for the same token pair, although they may belong to different protocols or have different fee tiers. In full generality, one might also allow trading through other intermediate tokens when routing. This raises additional challenges, such as determining the full universe of intermediate tokens. For this study, we restrict attention to the simplified same-token setting, which is nevertheless empirically relevant.

\medskip
\noindent \textbf{\emph{Output functions.}}
For pool $j$, we write the (fee–inclusive) \emph{output function}
\[
o_j(q_j,\omega_j,z)
\]
for the quantity of the output token obtained when supplying input $q_j\ge 0$ to
pool $j$ in state $\omega_j$ and swap direction $z\in\{0,1\}$.  We take
\[
z \;=\;
\begin{cases}
1, & \text{swap token 0 for token 1},\\
0, & \text{swap token 1 for token 0},
\end{cases}
\]
so that each pool possesses \emph{two} output mappings, one per direction.  The pool state $\omega_j$ captures all variables that influence execution on the pool, including, for example, the spot price on the pool, available liquidity and its distribution across prices, protocol fees, and any state variables used by hooks (e.g., recent volatility for dynamic fees).

We make several assumptions on these output functions to ensure well-posedness of the routing problem and to capture structural features of AMM liquidity curves observed in practice.

\begin{assumption}[Output–function regularity]\label{ass:ofn}
For every pool $j$ and direction $z\in\{0,1\}$, the mapping
$q\mapsto o_j(q,\omega_j,z)$ satisfies:
\begin{enumerate}
    \item \textbf{Nonnegativity:}
    $o_j(0,\omega_j,z)=0$ and $o_j(q,\omega_j,z)\ge 0$ for all $q\ge 0$.

    \item \textbf{Continuity and monotonicity:}
    $o_j(\cdot,\omega_j,z)$ is continuous and strictly increasing almost everywhere.\footnote{In practice, some pools have no available liquidity in certain price ranges, at which point $o_j$ can become locally constant. Such edge cases can be easily handled and for clarity of exposition we will not consider them here.}

    \item \textbf{Differentiability:}
    $o_j(\cdot,\omega_j,z)$ is continuously differentiable except at finitely many points.

    \item \textbf{Strict concavity (diminishing returns):}
    $o_j(\cdot,\omega_j,z)$ is strictly concave. Economically, execution deteriorates with size: each additional unit of input yields strictly less output than the previous unit.

\end{enumerate}
\end{assumption}

\medskip
\noindent \textbf{\emph{Gas cost.}}
For pool $j$ in state $\omega_j$ and input $q_j\ge 0$, let
$g_j(q_j,\omega_j)$
denote the gas expenditure required to execute the swap.  Without loss of generality, we treat gas as paid and measured in a fixed type of token used for swapping in this pool (e.g., token 1).
Its effect on the routing problem is asymmetric across swap directions: (i) In a token~0 $\to$ token~1 swap, gas reduces net token~1 output (\emph{output-side deduction}).
(ii) In a token~1 $\to$ token~0 swap, gas reduces the effective input budget in token~1 (\emph{input-side adjustment}).

\medskip
\noindent \textbf{\emph{Deterministic routing formulation.}}
Given the most recently observed pool states $\{\omega_j\}_{j=0}^{n-1}$, we allocate a fixed input budget $Q>0$ across the candidate pools $\mathcal{J}=\{0,1,\ldots,n-1\}$ to maximize total output \emph{net of gas}:
\begin{equation}
\label{eq:det-max-output-full}
\begin{aligned}
\max_{q\in\mathbb{R}_+^{\,n}}\quad
& \sum_{j=0}^{n-1}\Big[\, o_j(q_j,\omega_j,z)
 - (1-z)\, g_j(q_j,\omega_j)\,\mathbf{1}\{q_j>0\}\,\Big] \\[3pt]
\text{s.t.}\quad
& \sum_{j=0}^{n-1}\Big[\,q_j + z\, g_j(q_j,\omega_j)\,\mathbf{1}\{q_j>0\}\,\Big] \; \le \; Q .
\end{aligned}
\end{equation}
Here $z\in\{0,1\}$ encodes the swap direction for pool $j$ and determines where gas enters:
when $z=1$, gas is paid in the \emph{input} token and therefore appears in the resource
constraint; when $z=0$, gas is paid in the \emph{output} token and therefore deducts from the
objective. 

\subsection{Optimal Routing Solution without Gas Adjustment}
\label{subsec:optimal-multi}
We begin with the routing problem \emph{without} gas costs, i.e.\ $g_j(\cdot,\cdot)\equiv 0$:
\begin{equation}
\label{eq:max-output-nogas}
\max_{q\in\mathbb{R}_+^{\,n}}
\;\sum_{j=0}^{n-1} o_j(q_j,\omega_j,z)
\quad\text{s.t.}\quad
\sum_{j=0}^{n-1} q_j \le Q .
\end{equation}
At a fixed snapshot we treat pool states as given and suppress them, writing
$o_j(q_j)\equiv o_j(q_j,\omega_j,z)$. By Assumption~\ref{ass:ofn} (strict monotonicity), any unallocated budget can always increase output, so the resource constraint binds at the optimum. The inequality constraint in \eqref{eq:max-output-nogas} can be replaced by $\sum_{j=0}^{n-1} q_j = Q$. This model provides the deterministic benchmark and serves as the inner optimizer used later when we incorporate gas adjustments.

\medskip
\noindent \textbf{\emph{Lagrangian and optimality conditions.}}
Introducing a multiplier $\lambda$ for the budget constraint and $\mu_j\ge 0$ for non-negativity, the Lagrangian equation is
\begin{equation}
\label{eq:Lagrangian}
\mathcal{L}(q,\lambda,\mu)
=\sum_{j=0}^{n-1} o_j(q_j)
+ \lambda\!\left(Q-\sum_{j=0}^{n-1} q_j\right)
+ \sum_{j=0}^{n-1} \mu_j q_j .
\end{equation}
By concavity of the objective and convexity of the feasible set, first-order KKT conditions are necessary and sufficient for optimality.  

We next define the \emph{marginal output} $m_j(q)\equiv \partial o_j(q)/\partial q$ and the associated \emph{marginal price} $\mathrm{mp}_j(q)\equiv 1/m_j(q)$, which could be interpreted as the instantaneous input required to obtain one additional unit of output from pool $j$ when the allocation is $q$. The KKT conditions imply
\begin{equation}
\mathrm{mp}_j(q_j^*)- \lambda^* + \mu_j^* = 0,\qquad
q_j^*\ge 0,\ \mu_j^*\ge 0,\ \mu_j^* q_j^* = 0,\qquad
\sum_{j=0}^{n-1} q_j^* = Q ,
\label{eq:KKT}
\end{equation}
which is equivalent to this \emph{marginal equalization} rule
\begin{equation}
\mathrm{mp}_j(q_j^*)=\lambda^*\ \text{if } q_j^*>0,\qquad
\le \lambda^*\ \text{if } q_j^*=0.
\label{eq:marginal-equalization}
\end{equation}

Thus all active pools share a common marginal price $\lambda^*$, while inactive pools have weakly higher marginal costs. In other words, input is allocated so that shifting a marginal unit across pools cannot increase total output.

\medskip
\noindent \textbf{\emph{Inverse marginal price formulation.}} 
It is convenient to encode non-negativity directly via the \emph{clipped inverse marginal cost function} 
\begin{equation}
\mathrm{mp}_j^{-1}(\lambda)\ \equiv\
\inf\{\,q\ge 0:\ \mathrm{mp}_j(q)\ge \lambda\,\}
\label{eq:psi}
\end{equation}
where $\mathrm{mp}_j^{-1}(\lambda) = 0$ when $\lambda$ is less than current spot price $\mathrm{mp}_j^{-1}(0)$. The KKT system \eqref{eq:KKT} is equivalent to the fixed point
\begin{equation}
q_j^*=\mathrm{mp}_j^{-1}(\lambda^*)\quad(j=0,\dots,n\!-\!1),
\qquad
\sum_{j=0}^{n-1}\mathrm{mp}_j^{-1}(\lambda^*)=Q .
\label{eq:fixed-point}
\end{equation}

Let $S(\lambda)\equiv \sum_{j=0}^{n-1}\mathrm{mp}_j^{-1}(\lambda)$.  
By Assumption~\ref{ass:ofn}, each $\mathrm{mp}_j^{-1}(\cdot)$ is strictly increasing and right–continuous, and $S$ inherits these properties whenever $\lambda$ is higher than the current price.  Therefore, for any feasible budget $0<Q<\sup_\lambda S(\lambda)$ \footnote{Here $\sup_\lambda S(\lambda)$ denotes the maximum total input that can be absorbed across all pools, i.e., the aggregate capacity of the system.}, there exists a unique $\lambda^\star$ with $S(\lambda^\star)=Q$,  yielding a unique optimal allocation via \eqref{eq:fixed-point}.

\medskip
\noindent \textbf{\emph{Searching for the solution\footnote{The only oracle required is evaluation of $\mathrm{mp}_j^{-1}(\lambda)$. In both CPMMs and concentrated-liquidity AMMs this can be computed to machine precision.}}}
\label{subsubsec:bisection}
Since $S$ is monotone, $\lambda^\star$ can be computed efficiently by bisection on $[\lambda_{\mathrm{low}},\lambda_{\mathrm{high}}]$ with $S(\lambda_{\mathrm{low}})\le Q \le S(\lambda_{\mathrm{high}})$. At termination we obtain $\widehat\lambda$ with allocation $q^\star=\mathrm{mp}^{-1}(\widehat\lambda)$, which satisfies  the marginal equalization condition \eqref{eq:marginal-equalization}. Pseudocode (bracketing, stopping rules, and the evaluation oracle for $\mathrm{mp}_j^{-1}$) is given in Appendix~\ref{app:algorithms}, Algorithm~\ref{alg:bisection}. Formal guarantees, including existence, uniqueness and convergence, are stated in Theorem~\ref{thm:bisection}, with full statements and proofs deferred to Appendix~\ref{app:algorithms}.

\subsection{Optimal Routing with Gas Adjustment}
\label{subsec:Gas Adjustment Model}

We extend the deterministic routing problem to incorporate gas under an \emph{activation} model. Unlike the previous subsection (where $g_j\equiv 0$), we assume: (i) gas is paid in token 1; (ii) gas is incurred only when a pool is \emph{activated}, i.e., $q_j>0$; and (iii) conditional on activation, the gas cost is quantity–independent\footnote{In practice, some AMMs exhibit size-dependent gas. In concentrated-liquidity designs (e.g., Uniswap v3), gas usage increases with the number of ticks crossed, making $g_j(q,\omega_j)$ piecewise constant and weakly increasing in $q$. For tractability, we abstract to a fixed per-activation cost, which captures the dominant overhead costs while preserving concavity of the inner allocation problem.}. Formally, for pool $j$ in state $\omega_j$,
\begin{equation}
g_j(q_j,\omega_j) = \begin{cases}
0, & q_j=0,\\
\bar g_j(\omega_j), & q_j>0
\end{cases}
\end{equation}
This activation specification captures the fixed EVM overhead of calling a pool and cleanly separates the binary activation decision ($q_j>0$) from the continuous allocation across activated pools, preserving concavity of the inner allocation.

\medskip
\noindent \textbf{\emph{Gas–aware Program.}}
We define the gas-aware routing problem as below:
\begin{equation}
\label{eq:gas-mip-unified-ind}
\begin{aligned}
\max_{q\in\mathbb R_+^{\,n}}\quad
& \sum_{j=0}^{n-1} o_j(q_j,\omega_j,z)\;-\;\sum_{j=0}^{n-1} z\,\bar g_j(\omega_j)\,\mathbf{1}\{q_j>0\}\\[2pt]
\text{s.t.}\quad
& \sum_{j=0}^{n-1}\Big(q_j + (1-z)\,\bar g_j(\omega_j)\,\mathbf{1}\{q_j>0\}\Big)\;=\;Q.
\end{aligned}
\end{equation}
The equality budget is without loss: under Assumption~\ref{ass:ofn} the planner exhausts the available input. Problem \eqref{eq:gas-mip-unified-ind} is a mixed–integer program due to the activation terms $\mathbf{1}\{q_j>0\}$ and such problems are, in general, NP-hard.

\medskip
\noindent \textbf{\emph{Exact solution by subset enumeration.}}
Let $S\subseteq\mathcal{J}$ denote the active set of pools, i.e., those with $q_j>0$.  
For any fixed $S\neq\varnothing$, the gas terms are constants, and the problem reduces to a concave inner allocation:  

\begin{equation}
\label{eq:inner-fixed-S}
\begin{aligned}
Y(S)\;=\;
\max_{\{q_j\}_{j\in S}}\quad
& \sum_{j\in S} \Big(o_j(q_j,\omega_j,z)\;-\;z\,\bar g_j(\omega_j)\Big) \\[4pt]
\text{s.t.}\quad
& \sum_{j\in S} q_j \;=\; Q_{\mathrm{swap}}(S),\qquad q_j\ge 0\ \ (j\in S),
\end{aligned}
\end{equation}
where the effective swap budget is
$Q_{\mathrm{swap}}(S)\;=\;Q-\sum_{j\in S}(1-z)\,\bar g_j(\omega_j).$
Feasibility requires $Q_{\mathrm{swap}}(S)>0$. For any feasible $S$, the optimizer of \eqref{eq:inner-fixed-S} is characterized by \emph{marginal price equalization} across $j\in S$ and can be computed via the bisection procedure in Algorithm~\ref{alg:bisection} (Section~\ref{subsec:optimal-multi}).  
The outer problem then selects
\[
S^\star\in\arg\max_{S\subseteq\mathcal{J},\,S\neq\varnothing,\ Q_{\mathrm{swap}}(S)>0} Y(S),
\qquad
q^\star=\{q_j^\star(S^\star)\}_{j\in S^\star}.
\]

Complete pseudocode for the outer enumeration is provided in Appendix~\ref{app:algorithms}, Algorithm~\ref{alg:gas-subset}. Correctness and complexity are established in Theorem~\ref{thm:gas-runtime-ind} (Appendix~\ref{app:algorithms}).

\section{Suboptimality Benchmarks}
\label{sec:subopt-benchmarks}

In this section, we formalize the best–in–class benchmarks used to evaluate a realized route observed in an on-chain transaction. All evaluations are taken at a fixed pool state $\omega=\{\omega_j\}_{j\in\mathcal{J}}$. In the subsequent analysis, we allow this state to vary in order to compare allocations under different market conditions. Let the realized allocation be $q^{r}=(q^{r}_0,\ldots,q^{r}_{n-1})$ with total input $Q=\sum_{j=0}^{n-1}q^{r}_j$. The \emph{activated set} of pools is denoted $\mathcal{A} \;\defeq\; \{\, j\in\mathcal{J} : q^{r}_j > 0 \,\}$.

\medskip
\noindent\textbf{\emph{Optimized route benchmarks.}}  We construct three benchmarks that differ only in \emph{scope} (eligible pools) and \emph{gas treatment}. In each case, the corresponding program is solved at state $\omega$ to obtain
a best–in–class allocation $(S^\star,q^\star)$, which is then compared against the realized route
under identical modeling assumptions.

\medskip
\noindent \emph{(i) Support–Constrained Optimum (SCO; activated set, no gas).}
SCO evaluates \emph{split quality} conditional on the realized activated set
$\mathcal{A}$ and ignores gas:
\begin{equation}
\label{eq:SCO-def}
\begin{aligned}
\max_{\{q_j\}_{j\in \mathcal{A}}}\quad
& \sum_{j\in \mathcal{A}} o_j(q_j,\omega_j,z)\\
\text{s.t.}\quad
& \sum_{j\in \mathcal{A}} q_j = Q,\qquad q_j\ge 0.
\end{aligned}
\end{equation}

\noindent \emph{(ii) Full–Venue Optimum (FVO; all pools, no gas).}
FVO expands the scope to all candidate pools $\mathcal{J}$ while still ignoring gas. The program is
identical to \eqref{eq:max-output-nogas}.

\noindent \emph{(iii) Gas–Aware Full–Venue Optimum (G–FVO; all pools, gas–adjusted).}
G–FVO augments FVO by incorporating gas costs via model \eqref{eq:gas-mip-unified-ind} in Section~\ref{subsec:Gas Adjustment Model}.

\section{Empirical Results}
\label{sec:empirical}
This section evaluates realized router performance relative to the optimal programs defined in Section~\ref{sec:routing} and the benchmarks in Section~\ref{sec:subopt-benchmarks}. The analysis addresses three questions: \emph{(i)} How do realized routes compare to the best allocations under alternative assumptions about the pool universe and gas costs? \emph{(ii)} How valuable is it to use the most recent pool state for routing? \emph{(iii)} Do certain routers appear to route in advance of on-chain state updates? 

\medskip
\emph{\textbf{Data description}}
Our baseline sample comprises $2.98$ million WETH-USDC swaps on Ethereum mainnet
(blocks $19{,}500{,}000$ to $23{,}000{,}000$), totaling \$120.42\, billion in USDC-equivalent input. 
Transactions are reconstructed by aggregating swaps across pools at the transaction level and filtered to drop counterflow (both directions on the same pair) and trades with less than \$100 on the USDC side. We evaluate routes across four Uniswap L1 pools (v2: 30\,bps; v3: 1, 5, and 30\,bps) and, for benchmarking, label five routers (Universal Router, CoWSwap, 1inch v4, 1inch v5, Odos v2), which together account for about 21\% of transactions and 5.6\% of volume. Routes are highly concentrated: on average, fewer than one-third of the four pools are activated per transaction and flow shares are extremely uneven (low entropy, HHI $\approx 1$), suggesting potential under-activation of venues that we quantify in the FVO/G-FVO comparisons below. Full data construction details and summary tables appear in Appendix~\ref{sec:data}.

\subsection{Performance Relative to Optimal Benchmarks}
\label{subsec:perf-vs-opt}

\begin{figure}[h]
    \centering
    \includegraphics[width=0.95\linewidth]{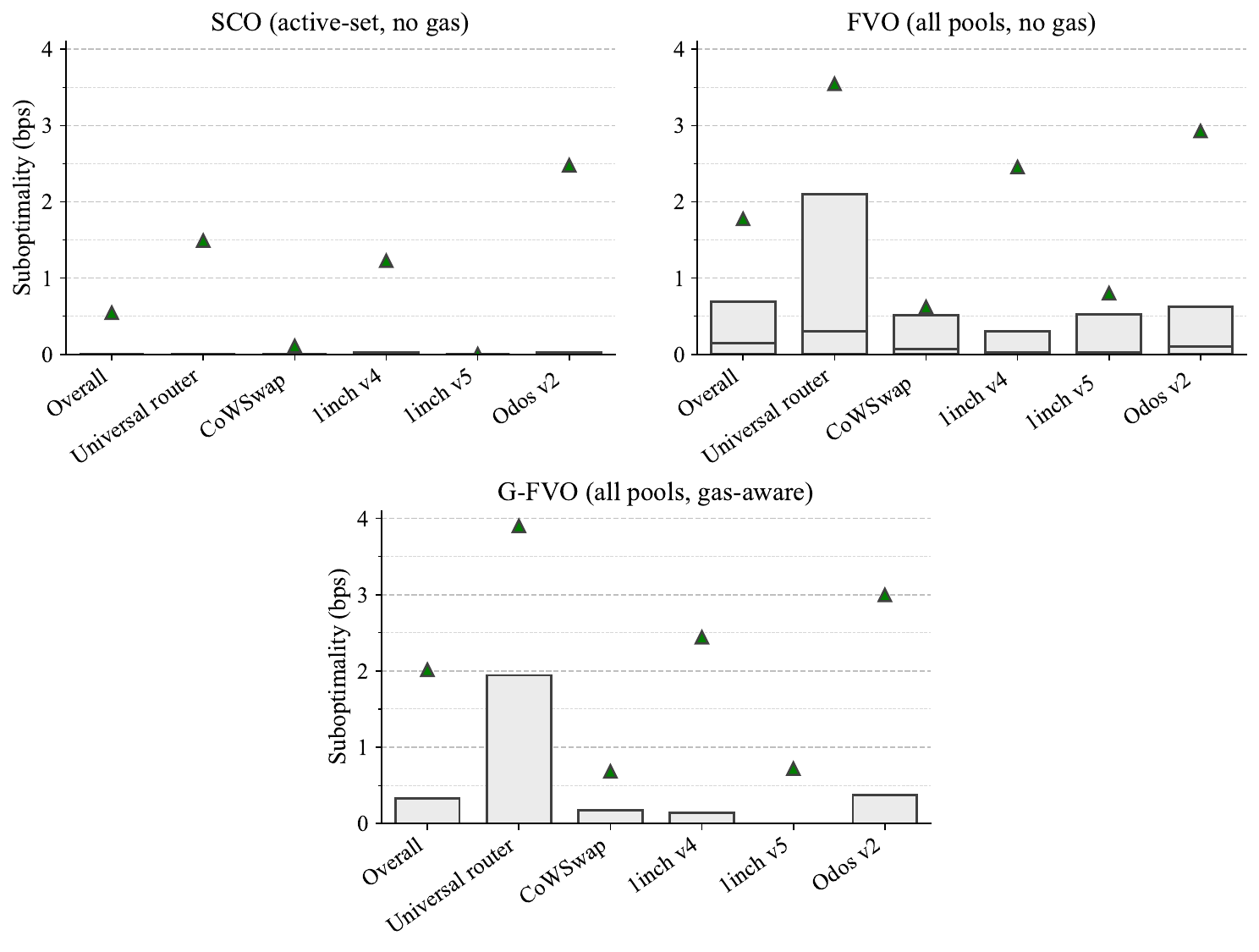}
    \caption{Suboptimality relative to different optima. The box plot indicates the 25th, 50th, and 75th percentiles, while the triangle indicates the mean of each sample.}
    \label{fig:subopt-router}
\end{figure}

We first compare realized routes with the three benchmarks developed in Section~\ref{sec:subopt-benchmarks}, all optimized and evaluated at execution states. Figure~\ref{fig:subopt-router} reports router–level box plots of suboptimality, defined as the proportional shortfall of the realized route relative to the benchmark (positive values indicate underperformance of the realized route):

\medskip
\emph{\textbf{Economic magnitudes.}}
Although Figure~\ref{fig:subopt-router} shows bp-level gaps, the implied dollar amounts are non-trivial, indicating the practical value of better activation and splitting across pools. Across all trades in our sample (\$120.42 billion; Table~\ref{tab:summary-sample}), the losses of realized routes amount to \$6.58 million relative to the SCO benchmark, \$21.39 million relative to the \emph{FVO} benchmark, and \$24.20 million relative to the \emph{G-FVO} benchmark. For the five labeled routers as a group, the same conversion yields losses of about \$0.77 million relative to SCO, \$1.77 million relative to FVO, and \$1.89 million relative to \emph{G-FVO}.

To put these magnitudes in perspective, one naturally asks: \emph{Where do such losses originate?} 

\emph{\textbf{Losses from mis–splitting within the activated set.}} 
Even conditional on the set of pools that were actually activated, realized splits are not fully efficient. Relative to \emph{SCO}, the mean shortfall is $0.05467$\,bps, which aggregates to approximately \$6.58\,million across all trades in our sample. This reflects pure within–set inefficiency: the router touched the right pools but did not allocate across them optimally.

\emph{\textbf{Losses from not activating additional pools.}} The scope for improvement through activating additional pools is also significant. As shown in Figure~\ref{fig:subopt-router}, moving from \emph{SCO} to \emph{FVO} shifts the suboptimality distributions upward markedly, and although \emph{G–FVO} attenuates the improvement once per–pool gas is charged, the effect remains large. Failure to activate additional pools accounts for the dominant share of inefficiency: even after incorporating per–pool gas, the aggregate loss is about \$24.27\,million. Hence, while mis–splitting is economically meaningful, the overwhelming driver of suboptimality is limited pool activation.

\begin{figure}[h]
    \centering
    \includegraphics[width=0.95\textwidth]{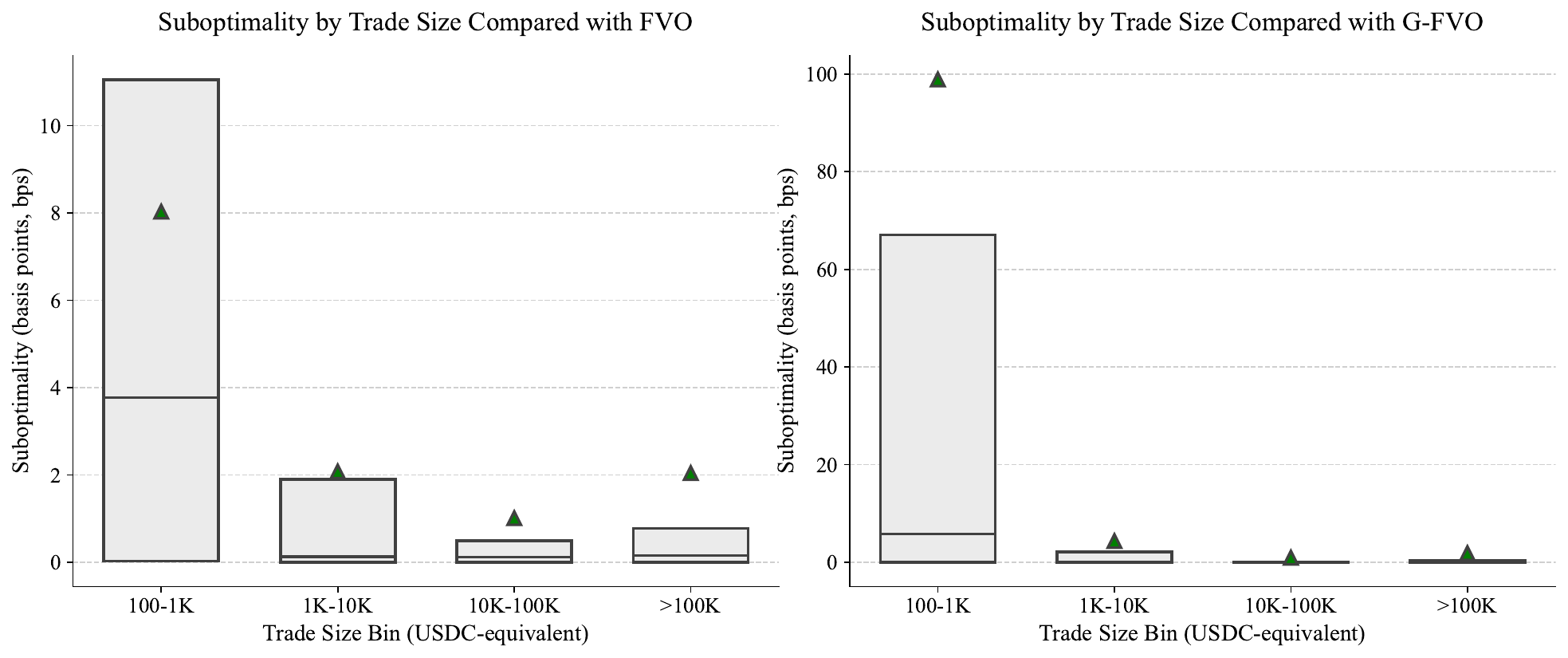}
    \caption{Suboptimality vs.\ trade size (relative to FVO).}
    \label{fig:subopt-size}
\end{figure}

\emph{\textbf{Heterogeneity by trade size.}}
Suboptimality is size–dependent. Figure~\ref{fig:subopt-size} displays the suboptimality of realized routes relative to FVO and G–FVO. In the FVO panel, the smallest bins (100–1K and 1K–10K USDC) have the widest interquartile ranges and visibly higher medians, whereas distributions tighten and center near zero for 10K–100K and $>$100K. Under G–FVO (right), small orders are even more dispersed with higher centers—consistent with fixed per–activation gas costs biting when notional is small—while large orders remain close to zero. Economically, although small trades are bps–inefficient, they contribute modestly to aggregate dollar loss, while large trades, despite low median bps, account for a meaningful share of total dollars forfeited due to their high notional.

\emph{\textbf{Impact of outliers.}}
A final observation is that suboptimality is heavily influenced by outliers. In both Figures \ref{fig:subopt-router} and \ref{fig:subopt-size}, the triangle mean markers lie well above the medians and often even exceed the upper quartile, while most boxes concentrate near zero. This pattern indicates long right tails: a small set of transactions contributes disproportionately to average losses, whereas the typical trade incurs little or no shortfall. Economically, these outliers account for much of the aggregate dollar gap reported above. Potential mechanisms include adversarial execution (e.g., sandwiching) or transient liquidity shocks, both of which can depress realized output relative to benchmark routes.

\subsection{Impact of Information Staleness}
\label{subsec:info-staleness} 
We measure the value of timely pool information via a controlled staleness design. For $N\in\{0,1,\ldots,10\}$, let $\omega^{N}=\{\omega^{N}_j\}_{j\in\mathcal{J}}$ be the state used \emph{both} to compute the benchmark and to evaluate outcomes:

(i) $N=0$ (\emph{oracle}): $\omega^{0}$ is the transaction’s execution-time state (the state at the trade’s actual execution position within its block).

(ii) $N\ge1$ (\emph{implementable}): $\omega^{N}$ is the bottom-of-block state $N$ blocks before execution—the freshest on-chain state a planner can deterministically observe before building the next block.

\begin{figure}[h]
    \centering
    \includegraphics[width=0.95\textwidth]{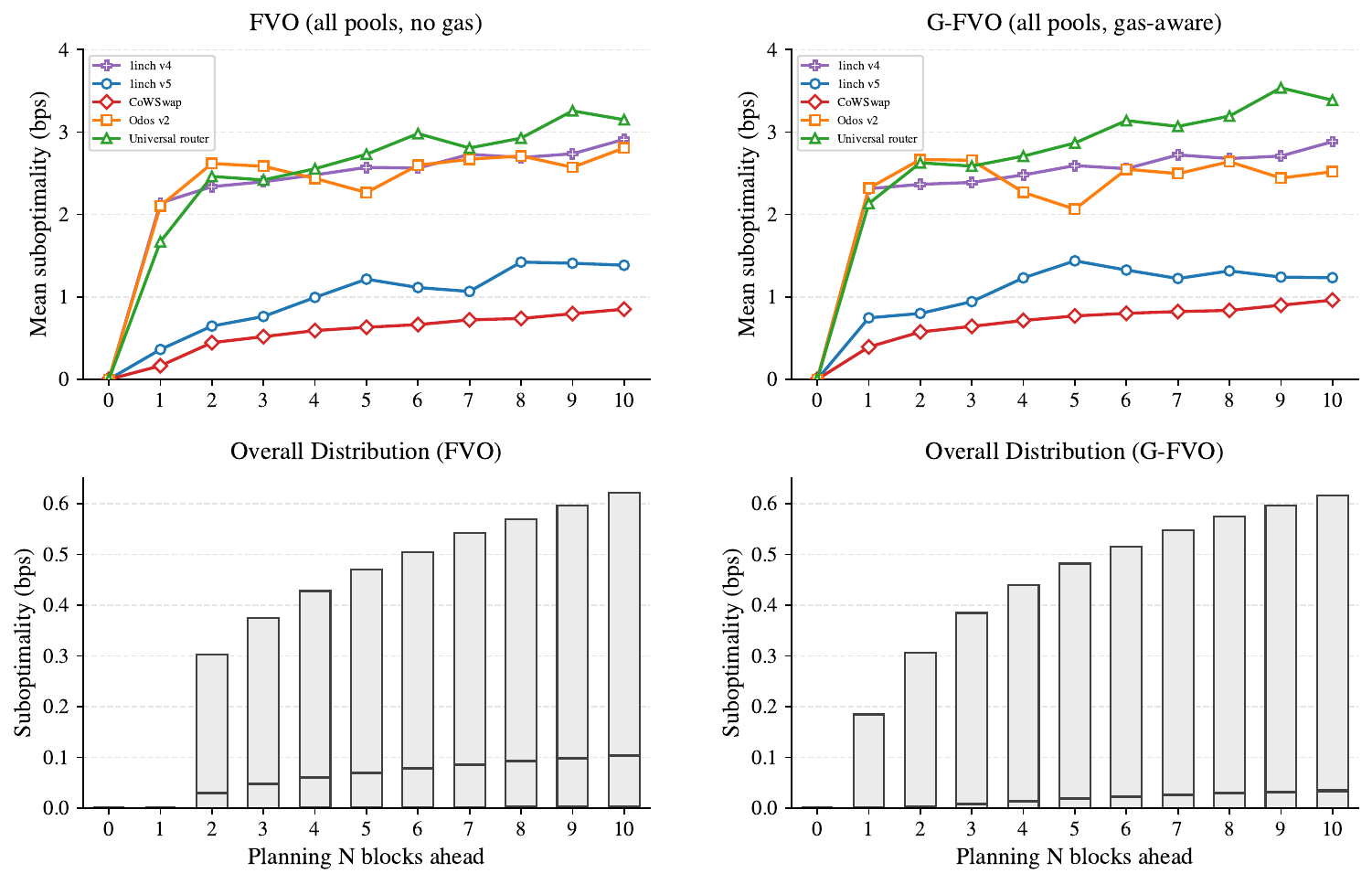}
    \caption{Effect of delayed state on performance (relative to FVO and G-FVO). 
    The gap between $N=0$ and $N=1$ quantifies the cost of one-block staleness 
    (oracle vs. implementable planning), while the monotone drift as $N$ increases 
    captures the growing loss from routing on progressively older snapshots.}
    \label{fig:metric2}
\end{figure}

For each $N$, we solve the optimized benchmark \footnote{We do not use the on-chain routes here; we consider only timing and total input quantity in Section~\ref{subsec:info-staleness}.} at $\omega^{N}$ and evaluate at the same $\omega^{N}$. Suboptimality measures the proportional shortfall of the optimized route using potentially outdated information at $\omega^{N}$, relative to the optimized route using up-to-date information available at execution. Figure~\ref{fig:metric2} plots suboptimality versus $N$ (router-level means as lines; overall interquartile box plots).

\emph{\textbf{One block of information is economically material.}} Moving from $N=0$ to $N=1$ (losing one block of state) almost doubles the aggregate loss. Using FVO, the mean shortfall rises from the oracle bound at $N=0$ to $N=1$ by $1.28739$ bps, which amounts to about \$15.5 million. Under G–FVO, the $N=0\to 1$ jump is $1.77525$ bps, roughly \$21.4 million. Thus, timely visibility of pool states delivers material value: even a single–block delay moves aggregate performance by eight figures, and additional delay continues to erode outcomes.

\textbf{Additional delay worsens performance, but with diminishing increments.} From $N=1$ to $N=2$, the incremental losses are much smaller, yet mean lines and interquartile ranges continue to drift upward with $N$, indicating cumulative degradation as information ages.

\textbf{Router heterogeneity and execution context.}
The mean curves for Universal Router and Odos v2 steepen more than those for CoWSwap, consistent with greater sensitivity to the most recent pool state. A plausible interpretation is execution context: users of Uniswap’s universal router often submit via public mempools, exposing them to reordering and sandwich risk \cite{Daian2020FlashBoys,Qin2022DarkForest}. By contrast, CoWSwap's flatter profile is consistent with solver-mediated batch auctions and private order flow handling that mitigate within-block state sensitivity \cite{CoWProtocol2023Whitepaper,CoWProtocol2023Docs}.

\subsection{Realized Routes vs.\ Planning-State Optima}
\label{subsec:realized-routes}

In this section we compare the realized routes to planning-state optima. We reuse the staleness design of Section~\ref{subsec:info-staleness} but consider two complementary evaluations.
\medskip

\textbf{(A) Evaluation at the planning state $\boldsymbol{\omega^N}.$} This isolates the role of timing mismatch versus allocation inefficiency and the measured gap answers: \textit{how much of the suboptimality reflects using pre-execution information, as opposed to suboptimal activation/splitting?}

\begin{figure}[h]
    \centering
    \includegraphics[width=0.98\textwidth]{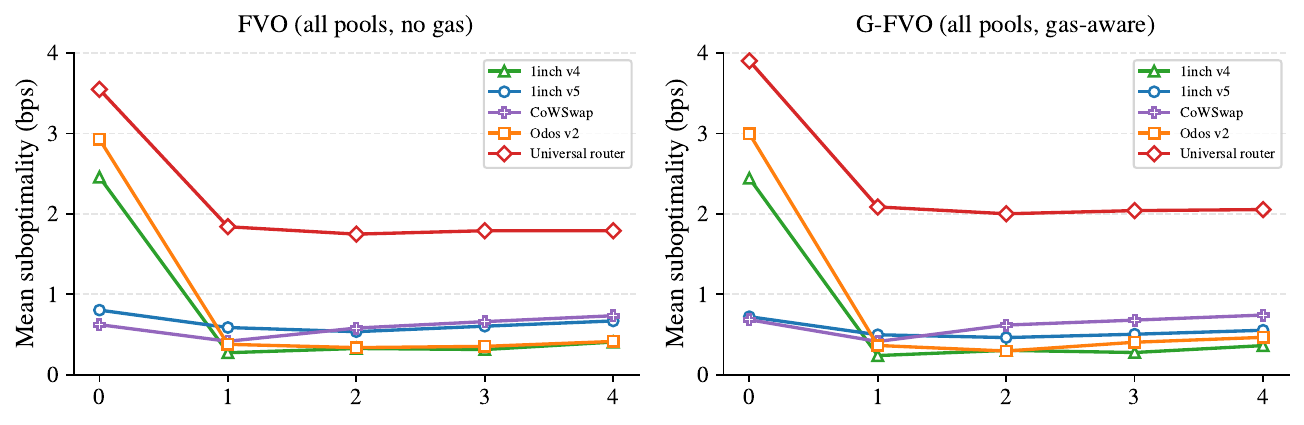}
    \caption{Suboptimality of realized routes against optima at $\omega^N$, evaluated also at $\omega^N$.}
    \label{fig:metric3}
\end{figure}

\textbf{Evaluated earlier, realized routes are less suboptimal.} Suboptimality declines monotonically with $N$: mean gaps fall as the benchmark is pushed further back in time. This pattern indicates that a material share of execution-time gaps stems from \emph{timing mismatch} (decisions keyed to pre-execution information), rather than purely to allocation inefficiency (suboptimal activation/splitting). 

\medskip
\textbf{(B) Implementable $\omega^{1}$ router vs. current routers}
We operationalize an implementable router that \emph{optimizes at the bottom of the previous block}, i.e., computes the FVO/G–FVO allocation at $\omega^{1}$ and submits the transaction in the next block. This uses the freshest on-chain state a planner can deterministically observe before building the next block. We then ask: \emph{at execution, how would an implementable router based on our algorithm perform relative to current routers?}

\begin{figure}[h]
    \centering
    \includegraphics[width=0.95\textwidth]{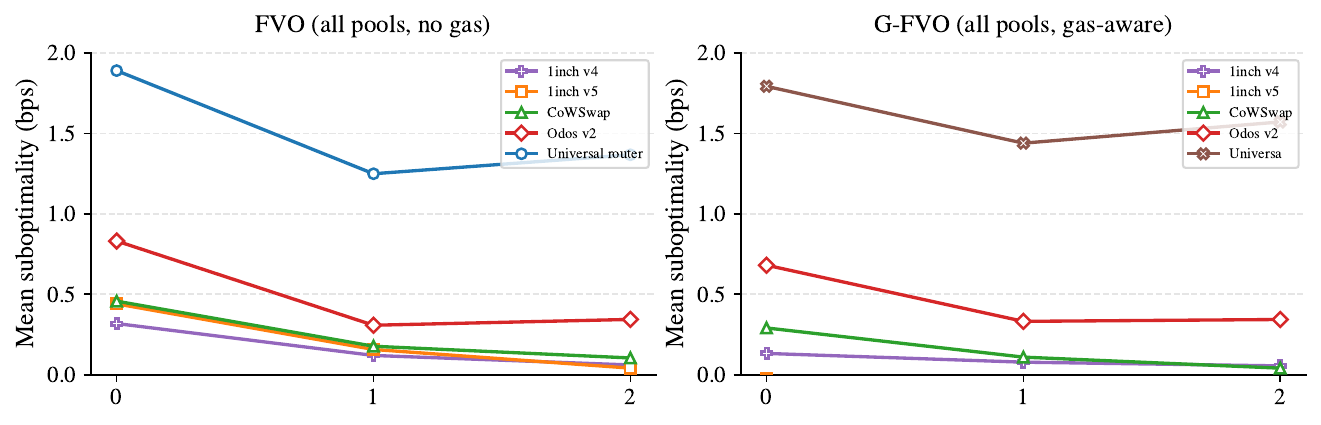}
    \caption{Suboptimality of realized routes against optima at $\omega^N$, evaluated at execution.}
    \label{fig:metric4}
\end{figure}

The execution–time comparisons indicate sizable gains. Figure~\ref{fig:metric4} shows that, when evaluated at execution, these $\omega^{1}$-based optima outperform realized routes across routers. Aggregating over the labeled routers’ total trade amount (Table~\ref{tab:summary-sample}), the implied gains amount to \$0.77 million for FVO$^1$ and \$0.66 million for G–FVO$^1$.

\subsection{Sandwiching Effect}
\label{subsec:sandwich-router}

We assess how sandwich attacks shift routing shortfalls at the \emph{router} level, keeping the evaluation identical to Section~\ref{subsec:perf-vs-opt}. For each pool in our universe (Section~\ref{sec:data}), we tag \emph{sandwiched} victim swaps using a within-block heuristic: a swap is labeled sandwiched on a pool if it is bracketed in the same block by two opposite-direction swaps attributable to the same counterparty cluster (front-run then back-run).  A transaction (its multi-pool combined route) is labeled sandwiched if at least one of the four pools flags as sandwiched. We then recompute suboptimality relative to the benchmarks in Section~\ref{sec:subopt-benchmarks}, optimizing and evaluating at \emph{execution-time} snapshots.

\begin{table}[htbp]
\centering
\footnotesize
\setlength{\tabcolsep}{3pt}
\begin{adjustbox}{max width=\linewidth} 
\begin{tabular}{lrrrrrr}
\toprule
\textbf{Router} &
\multicolumn{3}{c}{\textbf{Count}} &
\multicolumn{3}{c}{\textbf{Dollar value}} \\
\cmidrule(lr){2-4}\cmidrule(lr){5-7}
& \textbf{total} & \textbf{sandwiched} & \textbf{percentage}
& \textbf{total} & \textbf{sandwiched} & \textbf{percentage} \\
\midrule
Overall
& \(3.36\times10^{6}\) & \(1.03\times10^{4}\) & \(0.308\%\)
& \(1.21\times10^{11}\) & \(1.35\times10^{9}\) & \(1.12\%\) \\
Universal router
& \(5.85\times10^{5}\) & \(1.26\times10^{3}\) & \(0.215\%\)
& \(3.37\times10^{9}\)  & \(2.06\times10^{8}\) & \(6.12\%\) \\
CoWSwap
& \(4.29\times10^{4}\) & \(1.00\times10^{0}\) & \(0.00233\%\)
& \(9.90\times10^{8}\)  & \(4.53\times10^{4}\) & \(0.00458\%\) \\
1inch v4
& \(5.10\times10^{4}\) & \(7.97\times10^{2}\) & \(1.56\%\)
& \(1.83\times10^{9}\)  & \(1.45\times10^{8}\) & \(7.93\%\) \\
1inch v5
& \(8.73\times10^{4}\) & \(4.60\times10^{1}\) & \(0.0527\%\)
& \(3.84\times10^{8}\)  & \(2.61\times10^{6}\) & \(0.680\%\) \\
Odos v2
& \(3.39\times10^{3}\) & \(1.20\times10^{2}\) & \(3.54\%\)
& \(1.28\times10^{8}\)  & \(8.14\times10^{6}\) & \(6.34\%\) \\
\bottomrule
\end{tabular}
\end{adjustbox}
\caption{Sandwiched transactions summary by router}
\label{tab:router_sandwich_wide}
\end{table}

\noindent\textbf{Summary statistics.} The table summarizes the incidence of sandwiching by count and by USDC-equivalent volume across routers. It highlights pronounced cross-router heterogeneity: dollar exposure concentrates in Universal, 1inch v4, and Odos v2, whereas CoWSwap is negligible and 1inch v5 remains modest. This pattern aligns with the execution-context interpretation in Section~\ref{subsec:info-staleness}: publicly submitted flow (e.g., Universal/Odos) is more exposed to reordering and sandwich risk, while solver-mediated batch auctions and private order flow (CoWSwap) mitigate within-block state sensitivity.

\begin{figure}[H]
    \centering
    \includegraphics[width=0.9\textwidth]{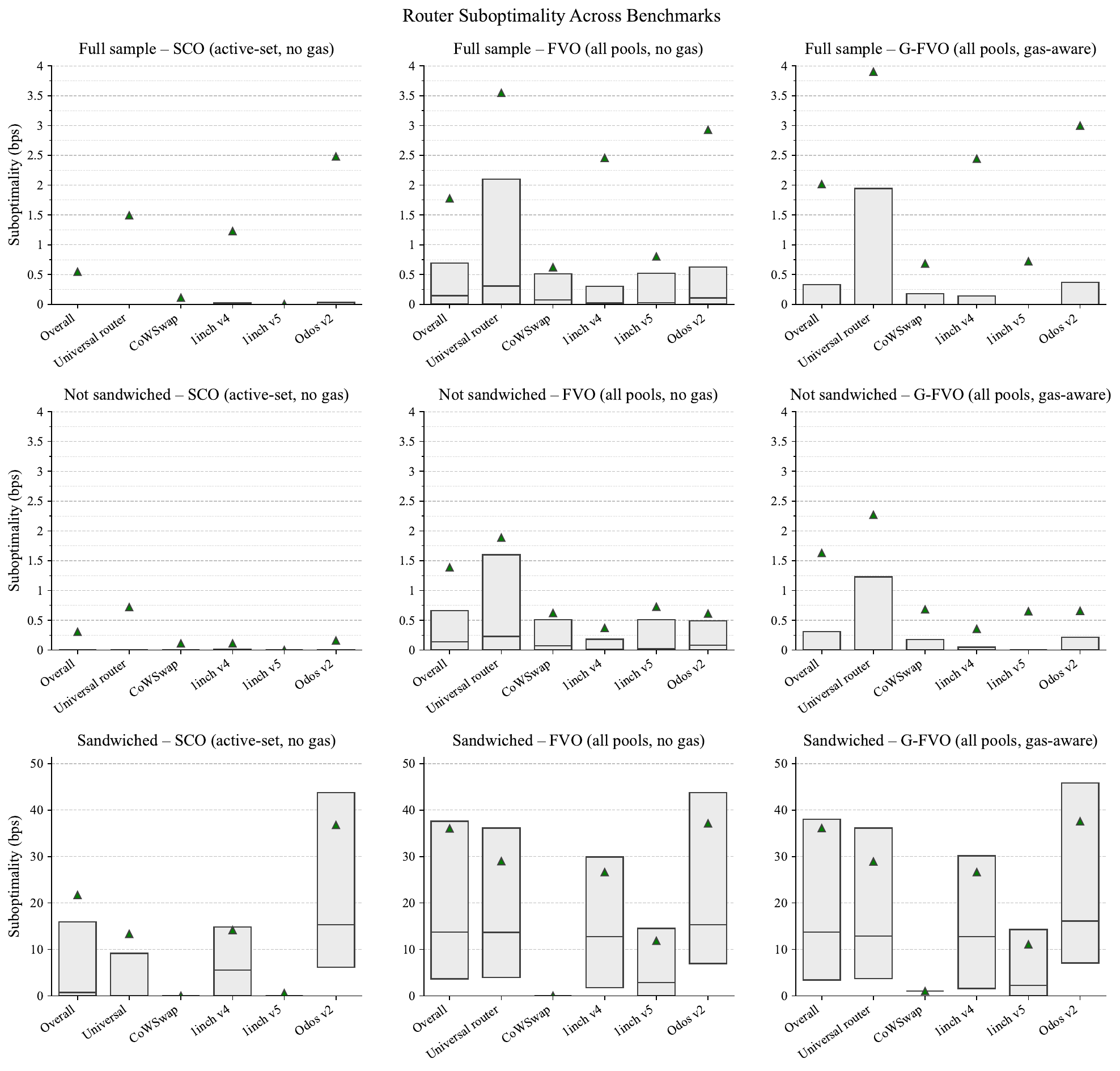}
    \caption{Router-level suboptimality relative to \emph{SCO}/ \emph{FVO}/ \emph{G-FVO}, evaluated at execution snapshots. Rows show the full sample (top), non-sandwiched subset (middle), and sandwiched subset (bottom). Boxes indicate the 25th/50th/75th percentiles; triangles indicate means.}
    \label{fig:sandwich-router-grid}
\end{figure}

\medskip
\noindent\textbf{Level shift and dispersion under sandwiching.}
Comparing the middle and bottom rows of Figure~\ref{fig:sandwich-router-grid} reveals a clear upward shift in both levels and dispersion once we restrict to sandwiched transactions. 
Non-sandwiched distributions cluster near zero in the low single-digit bps range, whereas the sandwiched panels expand to tens of bps with heavier right tails. 
This echoes Section~\ref{subsec:perf-vs-opt}: average shortfalls are driven by a minority of extreme observations, which are concentrated in the sandwiched set.

\medskip
\noindent\textbf{Router-level comparisons.}
In the non-sandwiched subset, CoWSwap exhibits the lowest medians and tightest IQRs across panels; Universal Router and Odos v2 show higher baselines; 1inch v4/v5 lie in between (with v5 modestly tighter than v4). 
Restricting to sandwiched transactions (bottom row) preserves this ordering but amplifies gaps: Universal and Odos experience the largest upward shifts and the heaviest right tails, 1inch rises moderately, and CoWSwap shifts the least. 
These cross-router patterns align with execution context: solver-mediated batch auctions and private order flow (e.g., CoWSwap) reduce within-block state sensitivity, while publicly submitted flow faces greater exposure to adversarial ordering.

\section{Conclusion}
\label{sec:conclusion}

We conduct a large-scale empirical audit of on-chain routing for AMMs, pairing a reproducible benchmark suite—\emph{SCO} (within-activation split quality), \emph{FVO} (full-venue access without gas), and \emph{G-FVO} (gas-aware activation)—with an exact and scalable optimizer. This framework provides a unified and implementable yardstick for evaluating realized routes.

Three findings emerge. \emph{First}, realized routes leave meaningful value on the table: average shortfalls are on the order of a few bps per trade and aggregate to tens of millions of dollars in our sample. Decomposition shows that insufficient pool activation dominates mis-splitting within the activated set, even after charging per-pool gas. \emph{Second}, information timeliness is crucial: moving from oracle execution snapshots to one-block stale planning materially worsens outcomes. \emph{Third}, losses are heterogeneous and heavy-tailed: most trades are near-optimal in bps, but a small set of extreme errors drives mean shortfalls. These tails are amplified in transactions flagged as sandwiched and differ systematically across routers, consistent with execution-context exposure to reordering and MEV.

Taken together, the results motivate router designs that (i) expand pool access while explicitly pricing activation gas, (ii) minimize timing mismatch via planning on the freshest implementable state, and (iii) reduce within-block state sensitivity through solver-mediated batch auctions, private order flow, or other MEV-mitigation mechanisms. Our benchmarks offer a practical harness for benchmarking such designs and for A/B testing policy choices (e.g., activation thresholds, venue whitelists, and timing rules) under identical market states.

Our study has limitations that point to fruitful extensions. We restrict attention to a single token pair on L1 and to same-pair pools, abstract from multi-hop routes and cross-domain execution, and model gas activation as a fixed per-call cost (ignoring tick-cross variability). Extending the benchmarks to multi-hop routing, richer gas models, additional venues and chains, and end-to-end mempool interactions (including private relay selection) is a natural next step. Distinguishing causal impacts of adversarial execution from confounding volatility, and stress-testing robustness to rapidly evolving AMM features (e.g., hooks), are additional directions.

\section*{Disclosure}
The second author is a research advisor for Paradigm and for Uniswap Labs. This research was supported by a grant from the Columbia Center for Digital Finance and Technologies.
    
\clearpage
\bibliographystyle{splncs04}
\bibliography{literature}

\appendix
\newpage

\section{Algorithms and Implementation Details}
\label{app:algorithms}

\setcounter{algorithm}{0}
\subsection{Bisection for the Lagrange Multiplier}
\begin{algorithm}[H]
\caption{Bisection for the Lagrange Multiplier $\lambda^\star$ (Support $\mathcal{J}$, Budget $Q$)}
\label{alg:bisection}
\begin{algorithmic}[1]
\REQUIRE Monotone maps $\{\mathrm{mp}_j^{-1}\}_{j=0}^{n-1}$; budget $Q>0$; 
bracket $[\lambda_{\mathrm{low}},\lambda_{\mathrm{high}}]$ with 
$S(\lambda_{\mathrm{low}})\le Q \le S(\lambda_{\mathrm{high}})$; 
tolerances $\varepsilon_\lambda,\varepsilon_S>0$; maximum iterations $K$.
\ENSURE Approximate multiplier $\widehat\lambda$ and allocation $q^\star$.
\STATE $k\!\gets\!0$
\WHILE{$k<K$}
  \STATE $\lambda\!\gets\!(\lambda_{\mathrm{low}}\!+\!\lambda_{\mathrm{high}})/2$;\quad $S(\lambda)\!\gets\!\sum_j\mathrm{mp}_j^{-1}(\lambda)$
  \IF{$|S(\lambda)-Q|\le\varepsilon_S$ \OR $\lambda_{\mathrm{high}}\!-\!\lambda_{\mathrm{low}}\le\varepsilon_\lambda$} 
      \STATE \textbf{break}
  \ELSIF{$S(\lambda)<Q$}
      \STATE $\lambda_{\mathrm{low}}\!\gets\!\lambda$
  \ELSE
      \STATE $\lambda_{\mathrm{high}}\!\gets\!\lambda$
  \ENDIF
  \STATE $k\!\gets\!k+1$
\ENDWHILE
\STATE $\widehat\lambda\!\gets\!(\lambda_{\mathrm{low}}\!+\!\lambda_{\mathrm{high}})/2$;\quad $q_j^\star\!\gets\!\mathrm{mp}_j^{-1}(\widehat\lambda)$
\STATE \textbf{return} $\widehat\lambda,\ q^\star$
\end{algorithmic}
\end{algorithm}

\begin{theorem}[Correctness and Complexity of Bisection]
\label{thm:bisection}
\begin{enumerate}
\item (\emph{Existence and uniqueness}) There exists a unique $\lambda^\star\in
[\lambda_{\mathrm{low}},\lambda_{\mathrm{high}}]$ such that $S(\lambda^\star)=Q$, and the allocation $q^\star_j=\mathrm{mp}_j^{-1}(\lambda^\star)$ is the unique optimizer of \eqref{eq:fixed-point}.
\item (\emph{Convergence in $\lambda$}) After $k$ iterations of Algorithm~\ref{alg:bisection},
the current bracket contains $\lambda^\star$ and has width
$\lambda_{\mathrm{high}}^{(k)}-\lambda_{\mathrm{low}}^{(k)}\le
(\lambda_{\mathrm{high}}^{(0)}-\lambda_{\mathrm{low}}^{(0)})/2^k$.
Hence to achieve $|\widehat\lambda-\lambda^\star|\le \varepsilon_\lambda$ it suffices to take
\[
k \ge \Big\lceil \log_2\!\big( (\lambda_{\mathrm{high}}^{(0)}-\lambda_{\mathrm{low}}^{(0)})/\varepsilon_\lambda \big) \Big\rceil .
\]

\item (\emph{Convergence in output space}) If, in addition, $S$ is $L$-Lipschitz on the bracket,
then $|S(\widehat\lambda)-Q|\le L\,(\lambda_{\mathrm{high}}^{(0)}-\lambda_{\mathrm{low}}^{(0)})/2^k$.
Therefore to reach $|S(\widehat\lambda)-Q|\le \varepsilon_S$ it suffices that
\[
k \ge \Big\lceil \log_2\!\big( L\,(\lambda_{\mathrm{high}}^{(0)}-\lambda_{\mathrm{low}}^{(0)})/\varepsilon_S \big) \Big\rceil .
\]

\end{enumerate}
\emph{Proof sketch.} Since $S$ is continuous and increasing with opposite-sign deviations
at the bracket endpoints, the intermediate value theorem yields a unique root of $S(\lambda)-Q$.
Each bisection step halves the bracket, establishing convergence in $\lambda$. Convergence in output space follows by Lipschitz continuity of $S$. The resulting $q^\star$ satisfies the resource constraint and the marginal price equalization \eqref{eq:marginal-equalization}.
\end{theorem}

\subsection{Subset Enumeration for Gas--Aware Routing}
\begin{algorithm}[H]
\caption{Subset Enumeration for Gas–Aware Routing}
\label{alg:gas-subset}
\begin{algorithmic}[1]
    \REQUIRE Pool set $\mathcal{J}$, budget $Q$, states $\{\omega_j\}$, directions $\{z\}$, gas $\{\bar g_j(\omega_j)\}$.
    \ENSURE Optimal active set $S^\star$ and allocation $q^\star$ for \eqref{eq:gas-mip-unified-ind}.
    \STATE Initialize $Y^\star \gets -\infty$, $S^\star \gets \varnothing$, $q^\star \gets \mathbf{0}$.
    \FORALL{$S \subseteq \mathcal{J},\; S \neq \varnothing$}
      \STATE $Q_{\mathrm{swap}}(S) \gets Q - \sum_{j \in S} (1-z)\,\bar g_j(\omega_j)$.
      \IF{$Q_{\mathrm{swap}}(S) \le 0$}
         \STATE \textbf{continue}
      \ENDIF
      \STATE Solve \eqref{eq:inner-fixed-S} on $S$ using Algorithm~\ref{alg:bisection}, obtaining $\{q_j^\star(S)\}_{j\in S}$ and objective $Y(S)$.
      \IF{$Y(S) > Y^\star$}
         \STATE $(Y^\star, S^\star, q^\star) \gets (Y(S), S, \{q_j^\star(S)\}_{j\in S})$
      \ENDIF
    \ENDFOR
    \STATE \textbf{return} $(S^\star, q^\star)$.
\end{algorithmic}
\end{algorithm}

\begin{theorem}[Correctness and Complexity of Subset Enumeration]
\label{thm:gas-runtime-ind}
Under Assumption~\ref{ass:ofn}, Algorithm~\ref{alg:gas-subset} returns a global maximizer of \eqref{eq:gas-mip-unified-ind}. Let $T_{\mathrm{inner}}(\varepsilon)$ denote the time to solve a single inner allocation \eqref{eq:inner-fixed-S} to tolerance $\varepsilon$ via bisection (Theorem~\ref{thm:bisection}), i.e.
\[
T_{\mathrm{inner}}(\varepsilon)=O\!\left(\log\frac{\lambda_{\mathrm{high}}-\lambda_{\mathrm{low}}}{\varepsilon}\right).
\]
Then the worst–case runtime is
$
O\!\Big(2^{|\mathcal{J}|}\,T_{\mathrm{inner}}(\varepsilon)\Big),
$
which is exponential in the number of candidate pools due to subset enumeration.

\emph{Proof sketch.} For fixed $S$, Theorem~\ref{thm:bisection} ensures a unique optimizer, attained via Algorithm~\ref{alg:gas-subset}. Exhaustive search across feasible $S$ then delivers the global optimum.
\end{theorem}

\section{Data Description}
\label{sec:data}

This section documents the dataset used in our empirical analysis: the universe of pools and routers, the sample period, transaction construction and filters, and descriptive statistics relevant for the gas–aware routing problem.

\noindent \textbf{\emph{Block range.}}
The baseline sample covers Ethereum mainnet blocks $19{,}500{,}000$ (Mar-23-2024 09:34:59 PM +UTC) to block $23{,}000{,}000$ (Jul-26-2025 01:22:11 AM +UTC).

\medskip
\noindent \textbf{\emph{Pools.}}
We study four Ethereum L1 Uniswap pools for the WETH–USDC pair: (i) the Uniswap v2 pool with a 30\,bps fee and (ii) three Uniswap v3 pools at fee tiers of $1$\,bp, $5$\,bps, and $30$\,bps. Each pool is treated as a distinct element $j\in\mathcal{J}=\{0,1,\ldots,n-1\}$ with its own output function $o_j(\cdot,\omega_j, z)$ as in Section~\ref{subsection:Routing Problem Formulation}.

\medskip
\noindent \textbf{\emph{Transaction construction and filters.}} Transactions are reconstructed at the transaction level by parsing AMM swap events from individual pools and aggregating them by transaction hash. We apply the following filters to obtain a clean routing sample:

(i) Counterflow exclusion. We drop transactions containing swaps in \emph{both} directions on the same pair (e.g., USDC$\to$WETH and WETH$\to$USDC). This is because the desired direction of trade is ambiguous.
  
(ii) Minimum trade amount. We drop transactions where the USDC side, summed across all pools used in the transaction, is less than $\$100$.

\medskip
\noindent \textbf{\emph{Routers.}} The sample includes \emph{all} transactions subject to the filters above. For benchmarking purposes, we also label trades executed by five widely used routing contracts: Uniswap \emph{Universal Router}, \emph{CoWSwap}, \emph{1inch v4}, \emph{1inch v5}, and \emph{Odos v2}. Labels are assigned by destination contract address in transaction logs.

\medskip
\noindent \textbf{\emph{Transaction counts and sizes.}} Table~\ref{tab:summary-sample} summarizes transaction counts, trade amount and average trade size. The core sample contains $2.98$ million transactions totaling \${120.42}\,billion in USDC–equivalent input. The five labeled routers jointly account for about $21\%$ of transactions and $5.6\%$ of trade amount.

\begin{table}[htb]
\centering
\setlength{\tabcolsep}{6pt}
\begin{tabular}{lrrr}
\toprule
\textbf{Router} & \textbf{Count} & \textbf{Trade Amount (USDC)} & \textbf{Avg. Trade Size (USDC)} \\
\midrule
Overall        & 2,978,019 & 1.2042$\times 10^{11}$ & 40,437 \\
Universal      &   456,046 & 3.3653$\times 10^{9}$  & 7,379 \\
CoWSwap        &    41,901 & 9.8510$\times 10^{8}$  & 23,510 \\
1inch v4       &    47,812 & 1.8241$\times 10^{9}$  & 38,151 \\
1inch v5       &    77,769 & 3.8342$\times 10^{8}$  & 4,930 \\
Odos v2        &     3,332 & 1.2841$\times 10^{8}$  & 38,540 \\
\bottomrule
\end{tabular}
\caption{Transaction counts and trade size for WETH–USDC swaps across the four Uniswap pools
(blocks $19{,}500{,}000$ to $23{,}000{,}000$).}
\label{tab:summary-sample}
\end{table}

\medskip
\noindent \textbf{\emph{Transaction concentration.}}
Since the gas model (Section~\ref{subsec:Gas Adjustment Model}) imposes a per–pool activation
cost, the degree to which order flow is concentrated across pools is empirically relevant. For
transaction $t$, let $q_{j,t}\ge 0$ be the input allocated to pool $j$ and define the normalized
share $x_{j,t} \;\equiv\; q_{j,t}\;/\;{\sum_{k\in\mathcal{J}} q_{k,t}}$. We measure concentration using three complementary statistics:

(i) \textbf{Fraction of pools activated:}
  $\kappa_t = \sum_{j\in\mathcal{J}} \mathbf{1}\{x_{j,t}>0\}/|\mathcal{J}|$.
  
(ii) \textbf{Entropy (dispersion):}
  $H_t \equiv -\sum_{j\in\mathcal{J}} x_{j,t}\log x_{j,t}$ (with $0\log 0\equiv 0$).
  
(iii) \textbf{Herfindahl–Hirschman Index (HHI):}
  $\mathrm{HHI}_t = \sum_{j\in\mathcal{J}} x_{j,t}^2 \in [1/|\mathcal{J}|,1]$.
  
\noindent We report transaction–weighted means of these statistics by router and provide trade-size–weighted means: 

\begin{table}[H]
\centering
\setlength{\tabcolsep}{3.5pt}
\begin{tabular}{lrrrrrr}
\toprule
\textbf{Metric} & \textbf{Overall} & \textbf{Universal} & \textbf{CoWSwap} & \textbf{1inch v4} & \textbf{1inch v5} & \textbf{Odos v2} \\
\midrule
Activation Rate & 0.283 & 0.296 & 0.324 & 0.380 & 0.250 & 0.355 \\
Entropy & 0.048 & 0.090 & 0.123 & 0.195 & 0.0004 & 0.175 \\
HHI     & 0.971 & 0.942 & 0.923 & 0.882 & 0.9997 & 0.892 \\
\bottomrule
\end{tabular}

\caption{Execution concentration metrics by router. Values are transaction–weighted means. Lower entropy and higher HHI indicate more concentrated execution.}
\label{tab:concentration}
\end{table}

On average, fewer than one–third of the four pools are activated per transaction. Entropy is close to zero, indicating that flow shares are highly unevenly distributed rather than split evenly across pools. Consistently, the Herfindahl–Hirschman Index (HHI) is near one, confirming that realized routes are almost always dominated by a single pool.

\end{document}